\documentclass[twocolumn,aps,superscriptaddress]{revtex4}

\usepackage{graphicx}
\usepackage{dcolumn}
\usepackage{bm}
\usepackage{amsmath,amssymb,amsfonts}
\usepackage{color}
\usepackage{braket}
\usepackage{mathtools}
\usepackage{siunitx}

\newcommand{\normord}[1]{:\mathrel{#1}:}

\begin{document}

\title{Nonlinear edge transport in a quantum Hall system}

\author{Hiroki Isobe}
\affiliation{Department of Physics, Kyushu University, Fukuoka 819-0395, Japan}
\affiliation{RIKEN Center for Emergent Matter Science (CEMS), Wako, Saitama 351-0198, Japan}

\author{Naoto Nagaosa}
\affiliation{RIKEN Center for Emergent Matter Science (CEMS), Wako, Saitama 351-0198, Japan}
\affiliation{Fundamental Quantum Science Program, TRIP Headquarters, RIKEN, Wako, Saitama 351-0198, Japan}

\begin{abstract}
Nonlinear transport phenomena in condensed matter reflect the geometric nature, quantum coherence, and many-body correlation of electronic states.  Electric currents in solids are classified into (i) Ohmic current, (ii) supercurrent, and (iii) geometric or topological current.  While the nonlinear current-voltage $(I$-$V)$ characteristics of the former two categories have been extensive research topics recently, those of the last category remains unexplored.  Among them, the quantum Hall current is a representative example.  
Realized in two-dimensional electronic systems under a strong magnetic field, the topological protection quantizes the Hall conductance in the unit of $e^2/h$ ($e$: elementary charge, $h$: Planck constant), of which the edge transport picture gives a good account. 
Here, we theoretically study the nonlinear $I$-$V_\text{H}$ characteristic of the edge transport up to third order in $V_\text{H}$.  We find that nonlinearity arises in the Hall response from electron-electron interaction between the counterpropagating edge channels with the nonlinear energy dispersions.  We also discuss possible experimental observations.
\end{abstract}

\maketitle

\section{Introduction}

Nonlinear electronic responses in solids, i.e., the nonlinear current-voltage ($I$-$V$) characteristics, are indispensable for the entire field of electronics, as exemplified by diodes and transistors \cite{transistor,Shockley}.  
Directional transport and nonreciprocal response are closely tied with the lack of inversion symmetry \cite{Tokura-Nagaosa}, whereas the nonlinearity is verily inherent in most solids as resistance, invoking Joule heating.  
In contrast, a superconducting current is perfectly dissipationless and hence there is no voltage drop $V$ up to a critical current $I_\text{c}$, where $V$ appears in a nonlinear manner.  
On the other hand, the superconducting diode effect \cite{Ono,Bauriedl,Daido,Yuan,JHe,Tanaka,Lu,JHe2} and the Josephson diode effect \cite{Hu,Misaki,Baumgartner,Ali,Pal,Trahms,Efetov} feature the asymmetry of $I_\text{c}$ depending on the current direction, revealing nonreciprocal phenomena in superconductors.

There is yet another category of electric current: the geometric or topological current.  An archetypal example is the polarization current, induced by a change in the electric polarization.  The modern theory of electric polarization \cite{King-Smith,Vanderbilt} formulates an electric polarization from the cumulative charge carried by the polarization current during an adiabatic process, which originates from the topological Berry curvature.  
Regarding nonlinear responses, the nonlinear Hall effect \cite{Sodemann,Ma,Mak,Du-review,Tiwari,Lai,Yao,Jack}, quantum frequency doubling and high-frequency rectification by skew scattering \cite{high-frequency,Du,Nandy,He1,He2,Ma-review} also rely on the Berry curvature, though they suffer from Joule heating.  

Dissipationless current highlights the class of the geometric current. 
The quantum Hall effect \cite{QHE,Klitzing,Ando_QHE,Wakabayashi} is a renowned manifestation of the geometric current grounded on the fundamental topological origin.  It descends to closely related phenomena such as the anomalous and spin Hall effects \cite{AHE,wavepacket,SHE}. 
Their prominent feature in common is that each has a characteristic quantized transport quantity protected by a symmetry. 
Lifting the symmetry protection ceases the quantization, which may induce a nonlinear response \cite{Bhalla,Wang}.

The integer quantum Hall effect realizes a quantized Hall conductance $G_\text{H} = \nu e^2/h$ ($\nu$: integer, $-e$: electron charge, $h=2\pi\hbar$: Planck constant) \cite{Laughlin,Thouless,Prange,Aoki,TKNN}.  It brought the concept of topology to the condensed matter physics as well as the application to a standard for electrical resistance \cite{CODATA,Jeckelmann,Poirier}.  
The integer $\nu$ is attributed to the Chern number, a topological invariant associated with Landau level wave functions in the bulk \cite{Kohmoto,Niu}.  
Therefore, it is robust against perturbations, e.g., disorder, and electron-electron and electron-phonon interactions \cite{TKNN,Niu,AHE,wavepacket}, and it persists up to a critical current, at which a large amount of current penetrates into the bulk and the longitudinal voltage drop appears \cite{Cage1,Kawaji,Tsemekhman,Nachtwei,Alexander-Webber}.

\begin{figure}[b]
\centering
\includegraphics[width=\hsize]{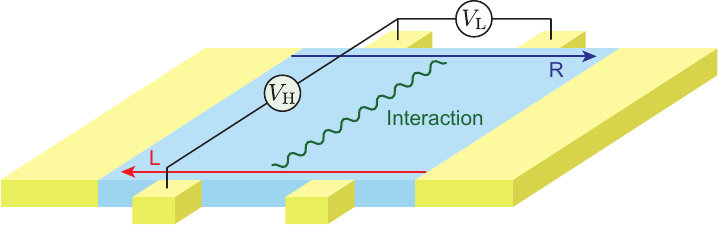}
\caption{Schematics for the nonlinear quantum Hall effect.
The edges of a quantum Hall system host chiral edge states according to the edge transport picture.  In a Hall bar setup, the two terminals (source and drain) induce electric current through the edge channels.  We attach voltage probes to measure the longitudinal voltage $V_\text{L}$ and the Hall voltage $V_\text{H}$.  
}
\label{fig:setup}
\end{figure}

The bulk-edge correspondence is another aspect of the topological electronic state \cite{QHE,Halperin}.  One-dimensional (1D) conduction channels appear along the edges of a two-dimensional sample, protected by the topological wave functions in the bulk.  One way to understand the quantized Hall conductance is to consider 1D transport through the edge conduction channels.  
In Fig.~\ref{fig:setup}, electrons flow through the upper and lower edge channels, driven by the Fermi level difference between the source and drain (left and right) terminals.  
Here, the upper channels accommodate right-moving electrons up to the Fermi level $E_\text{FR}$, and the lower channels do left-moving electrons up to $E_\text{FL}$.  Then, the Hall voltage is $V_\text{H} = (E_\text{FR}-E_\text{FL})/(-e)$ and the Hall resistance is $R_\text{H} (= G_\text{H}^{-1}) = V_\text{H}/I$ with the current $I$ flowing between the source and the drain.

\begin{figure}
\centering
\includegraphics[width=0.95\hsize]{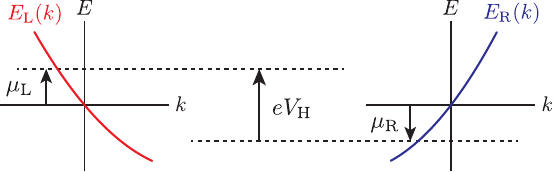}
\caption{Energy dispersions of the edge channels under a bias.  
The shifts of the Fermi levels in the terminals inject electrons to the edge channels.  The carrier densities vary from the equilibrium values, and the electrochemical potentials $\mu_\text{L/R}$ change, which reflect electron-electron interaction between the edge channels.   
The difference $\mu_\text{L} - \mu_\text{R}$ leads to the Hall voltage $V_\text{H}$. 
}
\label{fig:dispersion}
\end{figure}

In an experiment, we apply a finite current and measure the longitudinal and Hall voltages. 
The strict $I$-$V_\text{H}$ characteristics is essential for the exact quantization of the Hall conductance along with the vanishing longitudinal voltage. 
We suppose that each edge accommodates a single chiral channel and that the chemical potential of the left- and right-moving channels are $\mu_\text{L}$ and $\mu_\text{R}$, respectively (Fig.~\ref{fig:dispersion}).  
Then, the electric current \textit{in the absence of electron-electron interaction} is  
\begin{align}
\label{eq:linear}
I 
&= -\frac{e}{L} \sum_{k} [ - v_\text{L}(k) + v_\text{R}(k) ] \nonumber\\
&= \frac{e}{\hbar} \int_{0}^{k_\text{L}(\mu_\text{L})} \frac{dk}{2\pi} \frac{dE_\text{L}(k)}{dk} 
- \frac{e}{\hbar} \int_{0}^{k_\text{R}(\mu_\text{R})} \frac{dk}{2\pi} \frac{dE_\text{R}(k)}{dk} \nonumber\\
&= \frac{e^2}{h} V_\text{H}.  
\end{align}
Here $L$ is the length of the edge, and $v_\text{L(R)}(k) = dE_\text{L(R)}(k)/dk$ and $k_\text{L(R)}(\mu_\text{L(R)})$ are the group velocity and the Fermi wave vector of the left(right)-moving chiral channel.  
We consider a lattice model in the first line, making a transition to the thermodynamic limit in the second line.  
The $I$-$V_\text{H}$ characteristics is strictly linear without nonlinear terms regardless of the energy dispersion.   
When $\nu$ conduction channels reside along the edge, every channel equally contributes to electric current, and the quantized Hall conductance becomes $G_\text{H} = \nu e^2/h$.

However, the preceding discussion does not guarantee the linear $I$-$V_\text{H}$ relation \textit{in the presence of electron-electron interaction} though it does not alter the symmetry of a system. 
Oreg and Finkel’stein argued that the Coulomb interaction between the opposite edge channels with linear energy dispersions does not renormalize the linear Hall conductance \cite{Oreg1,Oreg2}.  
Nevertheless, a recent experiment observes nonlinear $I$-$V_\text{H}$ characteristics with a vanishing longitudinal voltage in monolayer graphene \cite{experiment}, and its theoretical interpretation is in high demand.  
Here, we investigate the possibility of the nonlinear response induced by electron-electron interaction in an integer quantum Hall system.  
We consider electron-electron interaction between the edge channels and reveal that the Hall conductance shows a nonlinear dependence on the applied current (voltage) when the energy dispersions along the edge are not exactly linear but have finite curvature.

\section{Results}

\subsection{Model}
\label{sec:model}

We consider the electronic response in a quantum Hall state based on the edge transport picture \cite{Halperin}.  
We adopt the Tomonaga--Luttinger liquid description for 1D systems \cite{Tomonaga,Luttinger} to calculate the Hall response.  It utilizes the bosonization technique, describing 1D fermions with a linear energy dispersion as harmonic bosons.  Its precisely harmonic nature implies the absence of nonlinear response without a nonlinear dispersion of the original 1D fermionic model.  
Therefore, we include a finite curvature of an 1D energy dispersions, i.e., a quadratic term with respect to wavenumber, which has been studied in Refs.~\cite{Haldane,Rozhkov1,Rozhkov2,Rozhkov3,Glazman2,Glazman3,Glazman4,Glazman_review}.

We suppose that the edge states contribute to the electronic transport and that they interact via the Coulomb interaction.  
For simplicity, we focus on the case with $\nu = 1$, while we discuss the extension to $\nu > 1$ later.  
We assume the nonlinear energy dispersions along the edge channels  
\begin{equation}
\label{eq:dispersion}
\begin{gathered}
E_\text{L}(k) = -v_\text{L} \hbar k + v_\text{L}' \hbar^2 k^2, \\
E_\text{R}(k) = v_\text{R} \hbar k + v_\text{R}' \hbar^2 k^2.  
\end{gathered}
\end{equation}
To analyze the model, we introduce the fermionic field operators $\psi_p(x)$ and $\psi_p^\dagger(x)$, satisfying the anticommutation relations 
\begin{equation}
\begin{gathered}
\label{eq:anticommutation}
\{ \psi_p(x), \psi_{p'}^\dagger(x') \} = \delta_{pp'} \delta(x-x'), \\
\{ \psi_p(x), \psi_{p'}(x') \} = \{ \psi_p^\dagger(x), \psi_{p'}^\dagger(x') \} = 0.  
\end{gathered}
\end{equation}
$p$ signifies the left ($+1$ or L) and right ($-1$ or R) movers.  
We then define the density operator   
\begin{equation}
\rho_p(x) = \normord{\psi_p^\dagger(x) \psi_p(x)}, 
\end{equation}
where $\normord{\cdots}$ describes the normal ordering with respect to the noninteracting ground state.  
The two edge channels interact via a two-body interaction $g(x-x')$, leading to the interaction energy 
\begin{equation}
E_\text{int} = \iint dx \, dx' \, g(x-x') \rho_\text{L}(x) \rho_\text{R}(x').  
\end{equation}

We suppose that the 1D electronic system has the length $L$ with the periodic boundary condition for analytic convenience.  It allows of the Fourier transformation 
\begin{gather}
\rho_{pq} = \int dx\, \rho_p(x) e^{-iqx}, \quad 
\rho_p(x) = \frac{1}{L} \sum_q \rho_{pq} e^{iqx}, 
\end{gather}
where $q = 2\pi m / L$ is the wavenumber  with an integer $m$.  
$\rho_{pq}$ obeys the commutation relation \cite{Mattis-Lieb}
\begin{equation}
\label{eq:commutation}
[ \rho_{pq}, \rho_{p' -q'} ] = p \eta_q \delta_{pp'} \delta_{qq'}, \quad \eta_q = \frac{qL}{2\pi}. 
\end{equation}
We require that the interaction range of $g(x)$ is shorter than the length $L$ to avoid complication by the periodic boundary condition, namely multiple interaction.  To wit, the interaction $g(x)$ should decay rapidly for $|x| \gg \lambda$, where the length scale $\lambda$ that characterizes the interaction range satisfies $\lambda \ll L$.

In a quantum Hall system, backscattering is exponentially suppressed with respect to the sample width $W$, i.e., the separation of the counterpropagating edge channels, as they are spatially distant with their wave functions localized near the opposite edges.  
We set $\hbar = 1$ hereafter unless otherwise noted.

The Tomonaga--Luttinger liquid description allows the mapping between 1D interacting and noninteracting models with linear energy dispersions.  We here extend the analysis by including the quadratic terms in the energy dispersions.  
The benefit of the mapping between the interacting and noninteracting models is that we can calculate the excited carrier density and the Fermi level shift from the equilibrium state of the interacting model via the noninteracting model, which we will see in the next section.  
The Fermi levels shift is observable as a voltage shift of the edge transport channel, which contributes to the Hall voltage.

The Hamiltonian for the edge transport model that we explained above is  
\begin{align}
H_\text{ex} 
&= \sum_p \int dx \left[ ip v_p \normord{\psi_p^\dagger \nabla \psi_p} + v'_p \normord{(\nabla \psi_p^\dagger) (\nabla \psi_p)} \right] \nonumber\\
&\quad + \iint dx\, dx'\, g(x-x') \rho_\text{L}(x) \rho_\text{R}(x') \nonumber\\
&\quad 
+ \sum_p ip g'_p \int dx\, [\normord{\psi_p^\dagger (\nabla \psi_p) - (\nabla \psi_p^\dagger) \psi_p}] \rho_{-p}  \nonumber\\
& \quad -\sum_p \mu_p \int dx\, \rho_p(x).    
\end{align}
$\mu_p$ is the chemical potential, which we require to be zero in the ground state.  We include it to \textit{externally} excite quasiparticles, which we will discuss in the next section.  
We add another interaction terms with the coefficient $g_p'$ to the Hamiltonian since they have the same scaling dimension as the energy dispersion curvature term \cite{Rozhkov1,Rozhkov2}.  We will later see that we should tune the coefficient $g'_p$ to make the model solvable.  
The bosonized form of the Hamiltonian is 
\begin{align}
\label{eq:Hex}
H_\text{ex} 
&= \sum_p \int dx \bigg( \pi v_p \normord{\rho_p^2} + \frac{4\pi^2}{3} v'_p \normord{\rho_p^3} \!\bigg) \nonumber\\
&\quad + \iint dx\, dx'\, g(x-x') \rho_\text{L}(x) \rho_\text{R}(x') \nonumber\\
&\quad + \sum_p 2\pi g'_p \int dx\, \normord{\rho_p^2} \rho_{-p} \nonumber\\
& \quad -\sum_p \mu_p \int dx\, \rho_p(x).    
\end{align}
In bosonization, the density operator $\rho_p(x)$ works as an elementary excitation.

In the following analysis, we make use of the result by Rozhkov \cite{Rozhkov1,Rozhkov2}.  
A central statement therein is that there exists a unitary transformation $U$ that acts on the bosonized form of the interacting Hamiltonian $H_\text{ex}$ to transform it into the noninteracting Hamiltonian $H_0$:
\begin{equation}
\label{eq:unitary-mapping}
U: H_\text{ex} \mapsto H_0.  
\end{equation}
The preceding works assumed that the left- and right-moving channels have the same velocities and curvatures ($v_\text{L} = v_\text{R}$, $v'_\text{L} = v'_\text{R}$).  
We extend the analysis by allowing them to have different values.  
While we present the explicit form of $U$ in Eq.~\eqref{eq:unitary}, the transformation $U$ acts on the Hamiltonian $H_\text{ex}$ as $U^\dagger H_\text{ex} U$, leading to the noninteracting Hamiltonian 
\begin{equation}
\label{eq:H0_bosonized}
H_0 = \sum_p \int dx\, \bigg( 
\pi \tilde{v}_p \normord{\tilde{\rho}_p^2} 
+ \frac{4\pi^2}{3} \tilde{v}'_p \normord{\tilde{\rho}_p^3}
- \tilde{\mu}_p \tilde{\rho}_p
\!\bigg).  
\end{equation}
$\tilde{\rho}_p(x)$ is another density operator, satisfying the same commutation relation as Eq.~\eqref{eq:commutation}.  
The density operator $\tilde{\rho}_p(x)$ induces the fermionic fields $\tilde{\psi}_p(x)$ and $\tilde{\psi}_p^\dagger(x)$, which results in the fermionic representation 
\begin{align}
\label{eq:H0}
H_0 &= \sum_p \int dx \Big[ ip \tilde{v}_p \normord{\tilde{\psi}_p^\dagger \nabla \tilde{\psi}_p} + \tilde{v}'_p \normord{(\nabla \tilde{\psi}_p^\dagger) (\nabla \tilde{\psi}_p)} \nonumber\\
&\qquad\qquad - \tilde{\mu}_p \normord{\tilde{\psi}_p^\dagger \tilde{\psi}_p} \!\Big].  
\end{align}

We recall that the interacting Hamiltonian $H_\text{ex}$ represents a physical system and that the noninteracting Hamiltonian $H_0$ is the fictitious solvable Hamiltonian.  
The parameters in the two models are related by  
\begin{subequations}
\label{eq:mapping}
\begin{gather}
\label{eq:v_modified}
\tilde{v}_p = v_p + \frac{2v'_p}{v_\text{L} + v_\text{R}} g_0 n_{-p} + O(g_0^2), \\
\tilde{v}'_p = v'_p + O(g_0^2), \\
\label{eq:mu_map}
\tilde{\mu}_p = \mu_p, 
\end{gather}
\end{subequations}
where we treat $g_0 = \int dx \, g(x)$ as a perturbative coupling constant.  The additional interaction parameter should obey
\begin{equation}
\label{eq:constraint}
g'_p = \frac{v'_p}{v_\text{L} + v_\text{R}} g_0 + O(g_0^2), 
\end{equation}
which completes the mapping between the interacting and noninteracting Hamiltonians and thus ensures the solvability of $H_\text{ex}$.  
We present the derivation of the mapping in the Materials and Methods.

\subsection{Carrier density}
\label{sec:carrier-density}

In the previous section, we observed the relation between $H_0$ and $H_\text{ex}$.  While we view $H_\text{ex}$ as a physical Hamiltonian, the mapping to the noninteracting model allows us to calculate the number of quasiparticle excitations relative to the ground state.  
The chemical potential $\mu_p$ is the externally controllable parameter that determines the number of electrons in the system connected to electron reservoirs.

From the noninteracting Hamiltonian Eq.~\eqref{eq:H0} along with Eq.~\eqref{eq:mu_map}, we obtain the relation between the chemical potential $\mu_p$ and the Fermi wavevector $k_{\text{F}p}$ relative to the ground state as 
\begin{equation}
\mu_p = -p \tilde{v}_p k_{\text{F}p} + \tilde{v}'_p k_{\text{F}p}^2.   
\end{equation}
In the thermodynamic limit, the Fermi wavevector directly gives the carrier density at zero temperature
\begin{align}
\label{eq:density}
n_p 
&= \frac{1}{2\pi} (-p k_{\text{F}p}) 
= \frac{1}{4\pi \tilde{v}'_p} \left( \sqrt{\tilde{v}_p^2 + 4\tilde{v}'_p \mu_p} - \tilde{v}_p \right) \nonumber\\
&= \frac{1}{2\pi} \left( \frac{1}{\tilde{v}_p} \mu_p - \frac{\tilde{v}'_p}{\tilde{v}_p^3} \mu_p^2 + \frac{2\tilde{v}_p^{\prime 2}}{\tilde{v}_p^5} \mu_p^3 \right) + O(\mu_p^4), 
\end{align}
where we identify the carrier density $n_p$ as the average density of quasiparticle excitations 
\begin{equation}
n_p = \frac{\langle N_p \rangle}{L}.  
\end{equation}
Here, we denote the quantum-mechanical average as $\langle \cdots \rangle$.  

In an experiment, we can measure the chemical potential $\mu_p$ with a voltmeter or a potentiometer.  When only the electrostatic potential $\Phi_p$ contributes to the chemical potential, one may think that $\mu_p$ and $-e\Phi_p$ are equal.  
However, the equality does not always hold in the presence of interaction.  
We follow the argument by Kawabata \cite{Kawabata} to self-consistently evaluate the chemical potential including electron-electron interaction.  

The self-consistent field method has a similarity to a mean-field approach.  We regard the chemical potential per electron as an electron single-particle energy.  When we neglect density-density correlations and fluctuations, the interaction modifies the chemical potential at the mean-field level as 
\begin{equation}
\label{eq:self-consistent}
\mu_p + g_0 n_{-p} + 4\pi g'_{-p} n_p n_{-p} + 2\pi g'_{-p} n_{-p}^2 = \bar{\mu}_p.  
\end{equation}
Here $\bar{\mu}_p$ is the self-consistent chemical potential, which is physically observable, and we attribute the measurable voltage $\Phi_p$ to it via 
\begin{equation}
\label{eq:voltage}
\bar{\mu}_p = -e\Phi_p.  
\end{equation}
From Eqs.~\eqref{eq:density} and \eqref{eq:self-consistent} with Eqs.~\eqref{eq:mapping} and \eqref{eq:constraint}, we find the carrier density as a function of the measurable chemical potential as 
\begin{align}
\label{eq:n-mu}
n_p &= \frac{1}{2\pi} \bigg( 
\frac{1}{v_p} \bar{\mu}_p - \frac{v'_p}{v_p^3} \bar{\mu}_p^2 + \frac{2 v_p^{\prime 2}}{v_p^5} \bar{\mu}_p^3
\bigg) \nonumber\\
&\quad + \frac{g_0}{4\pi^2} \bigg[ 
- \frac{\bar{\mu}_{-p}}{v_\text{L} v_\text{R}} 
+ \frac{v'_{-p} \bar{\mu}_{-p}^2}{v_{-p}^3 (v_\text{L} + v_\text{R})} 
- \frac{2 v_{-p}^{\prime 2} \bar{\mu}_{-p}^3}{v_{-p}^5 (v_\text{L} + v_\text{R})} \nonumber\\
&\qquad 
+ \frac{2 v'_p}{v_p^2 (v_\text{L} + v_\text{R})} \left( \frac{1}{v_p} - \frac{1}{v_{-p}} \right) \bar{\mu}_p \bar{\mu}_{-p} 
\nonumber\\
&\qquad 
- \frac{6 v_p^{\prime 2}}{v_p^4 (v_\text{L} + v_\text{R})} \left( \frac{1}{v_p} - \frac{1}{v_{-p}} \right) \bar{\mu}_p^2 \bar{\mu}_{-p} \nonumber\\
&\qquad 
+ \frac{2 v'_p v'_{-p}}{v_p^2 v_{-p}^3 (v_\text{L} + v_\text{R})} \bar{\mu}_p \bar{\mu}_{-p}^2
\bigg] 
+ O(g_0^2, \bar{\mu}_p^4).  
\end{align}

\subsection{Current response}

From the equation of continuity, we derive the mean current operator as \cite{Haldane}
\begin{align}
\label{eq:current-operator}
J_p 
&= -e \lim_{q \to 0} \frac{1}{q} [H_\text{ex}, \rho_{pq}] \nonumber\\
&= pe v_p N_p + 2\pi pe v'_p \frac{1}{L} \sum_{q'} \rho_{pq'} \rho_{p-q'} \nonumber\\
&\quad + pe \frac{g_0}{2\pi} N_{-p} + 2pe g'_p \frac{1}{L} \sum_{q'} \rho_{pq'} \rho_{-p-q'} \nonumber\\
&\quad + pe g'_{-p} \frac{1}{L} \sum_{q'} \rho_{-p q'} \rho_{-p -q'}.  
\end{align}
When we neglect the density-density correlation, the current response normalized by the system length becomes 
\begin{align}
&\quad\ I_p 
= \frac{\langle J_p \rangle}{L} \nonumber\\
&= pe \left( v_p n_p + 2\pi v'_p n_p^2 
+ \frac{g_0}{2\pi} n_{-p} + 2 g'_p n_p n_{-p} + g'_{-p} n_{-p}^2 \right).  
\end{align}
The total current response $I$ is the sum of the current carried by the left- and right-moving channels localized along the opposite edges.  
From Eqs.~\eqref{eq:voltage} and \eqref{eq:n-mu}, we obtain the current response as a function of the voltage shift $\Phi_p$ from the equilibrium on each edge: 
\begin{align}
\label{eq:current-response}
I 
&= \sum_p I_p \nonumber\\
&= \frac{e^2}{h} (\Phi_\text{R} - \Phi_\text{L}) \nonumber\\
&\quad + \frac{2 g_0 e^3 (v'_\text{L} - v'_\text{R})}{h^2 (v_\text{L} + v_\text{R}) v_\text{L} v_\text{R}} \nonumber\\
&\qquad \times \bigg(\! -\Phi_\text{L} \Phi_\text{R} 
- \frac{ev'_\text{L}}{v_\text{L}^2} \Phi_\text{L}^2 \Phi_\text{R} - \frac{ev'_\text{R}}{v_\text{R}^2} \Phi_\text{L} \Phi_\text{R}^2 \!\bigg) \nonumber\\
&\quad + O(g_0^2, \Phi_p^4).  
\end{align}
Here, we have recovered the Planck constant $h$, by substituting $v_p$ and $v'_p$ with $\hbar v_p$ and $\hbar^2 v'_p$, respectively.

The result indicates the conditions for nonlinear current response: 
(i) electron-electron interaction between the counterpropagating edge channels and 
(ii) different curvatures of the energy dispersions for the left- and right-moving channels.  
We note that the uniform Fourier component of the interaction $g_0$ contributes to the nonlinear response, which is natural considering momentum conservation; see also the Discussion.    
Since the nonlinear terms arise from electron-electron interaction, they vanish when either $\Phi_p$ is zero, where there is no quasiparticle in either edge channel.

One may find Eq.~\eqref{eq:current-response} is not gauge invariant in that a global shift of the voltages $\Phi_\text{L}$ and $\Phi_\text{R}$ would induce current.  
The gauge invariance is rooted in the particle number conservation.  
The problem arises from the normal ordering in the description of the 1D model.  
The normal ordering with respect to the noninteracting ground state deals with the divergences from infinite occupied states by subtraction, which does not guarantee the particle number conservation. 
The subtraction of divergences from occupied states is actually related to the commutation relation of the density operator Eq.~\eqref{eq:commutation}, as discussed by Mattis and Lieb \cite{Mattis-Lieb}.   
Concurrently, we should designate the ground state in the beginning of the analysis and the present result does not accept the global voltage shift as a consequence.  
In practice, we should identify the model parameters in the beginning of the analysis, and then we can introduce finite voltage shifts $\Phi_p$ to calculate current response.

\section{Discussion}
\label{sec:discussions}

We observe from Eq.~\eqref{eq:current-response} that the linear Hall conductance remains $e^2/h$, not renormalized by electron-electron interaction in the system.  
The quantization of the Hall conductivity is discussed also from Landau levels \cite{Kaplan} and Galilean invariance \cite{Hansson}. 
A 1D interacting quantum wire reveals a similar behavior that the longitudinal conductance is $e^2/h$ (per electron transport channel) regardless of interaction inside the observed system \cite{Tarucha}.  This behavior is explained for a finite size interacting system with noninteracting leads attached \cite{Maslov,Ponomarenko,Safi1,Safi2,Safi3} and without leads \cite{Oreg1,Oreg2,Kawabata,Shimizu}.

It is worth considering the nonlinear response from the viewpoint of thermodynamics \cite{Shimizu}.  When a system is in the thermodynamic equilibrium, the chemical potential of each edge channels is 
\begin{equation}
\label{eq:chemical_potential}
\bar{\mu}_p = \frac{\partial E_p}{\partial N_p}, 
\end{equation}
where $E_p$ is the internal energy of the edge channel $p$.  
On the other hand, the local current operator $J_p(x)$ in Eq.~\eqref{eq:current-operator} is equivalent to the functional derivative of the Hamiltonian with respect to the density $\rho_p(x)$ \cite{Oreg1,Oreg2}: 
\begin{equation}
J_p(x) = \frac{pe}{2\pi} \frac{\delta H_\text{ex}}{\delta \rho_p(x)}, 
\end{equation}
where we use the bosonized form of $H_\text{ex}$ here.  
Therefore, we immediately find a linear relation between the chemical potential $\mu_p$ and the uniform component of $J_p$, and hence nonlinear current response does not appear.  
The nonlinear response appearing in Eq.~\eqref{eq:current-response} seemingly violate the relation.  However, the thermodynamic relation Eq.~\eqref{eq:chemical_potential} may not be valid in the presence of a long-range interaction because it violates the additivity, meaning that the thermodynamic properties will be different if we divide the system into two subsystems.  
In the present case, electron-electron interaction couples the two edge channels, and the chemical potential of one edge channel affects the other with finite interaction $g_0 \neq 0$, as we can see from Eq.~\eqref{eq:n-mu}.  
We also remark on the boundary condition.  While we have adopted the periodic boundary condition to avoid boundary effects, an open boundary or a contact to a terminal would cause additional effects, which are not considered the present analysis.

It is also meaningful to argue the relation to Laughlin's argument \cite{Laughlin}, where the quantization of the integer Hall effect is attributed to the conservation of the electron number in an \textit{isolated} quantum Hall system.  
The gauge invariance imposed by electron number conservation concludes a charge pump by an adiabatic insertion of a magnetic flux quantum.  
It transports an integer number of electrons from delocalized states along one edge to the other through the substantially localized bulk states \cite{Halperin}.  The adiabatic time evolution does not induce excitations across the energy gap and the electric field is infinitesimally small throughout the time evolution.  
As a result, we can relate the number of electrons and the magnetic flux quantum to find the quantized Hall conductance.  
The linear $I$-$V_\text{H}$ characteristics thus hold between the current $I$ flowing through the bulk and the voltage difference $V_\text{H}$ between the two edges.  The Hall voltage $V_\text{H}$ originates from the difference of the chemical potentials in the edge channels caused by the electron transfer; in other words, it counts the number of electrons carried by the flux insertion.  
In our present discussion about the nonlinear response, on the other hand, we attach electron reservoirs to the edge channels to shift the chemical potentials.  As a result, the electron number is not conserved within each quantum Hall channel.  
Electrons injected to an edge channel from a reservoir rapidly thermalize to the chemical potential of the edge and do not carry information of the bulk.  
The external contributions to the chemical potentials of the edge channels and hence the voltage difference are essential for the nonlinear response.  

In the present analysis, we adopt the edge transport picture to analyze the nonlinear response of a quantum Hall system.  The underlying assumption of the discussion is that the localized states in the bulk do not contribute to electric transport.  The Hall voltage should not be too large to invoke a current path through the bulk or a transition between the Landau levels.  
An appearance of the longitudinal voltage drop is indicative of the breakdown of the quantum Hall state \cite{Cage1,Kawaji,Tsemekhman,Nachtwei,Alexander-Webber}.  
We also note that the energy dispersions [Eq.~\eqref{eq:dispersion}] contain terms to order $k^2$.  We should restrict a large chemical potential shift that changes the sign of the velocity, namely the chirality of the edge channel.

As we have discussed, electron-electron interaction and the curvatures of the energy dispersions along the edges induce the nonlinear response from the quantum Hall edge channels.  
Leaving the uncertainty of the energy dispersions, which reflect the details near the edges, we focus here on electron-electron interaction.  
In the present setup (Fig.~\ref{fig:setup}), there are two parallel one-dimensional edges, characterized by the length $L$ and the separation (width) $W$.  
Regarding the length of the edge channels, we have required the interaction range $\lambda$ to be shorter than $L$.  
As long as the condition is satisfied, the length $L$ does not cause a substantial difference according to the definition $g_0 = \int dx \, g(x)$.  
On the other hand, the width dependence is more involved.  If we suppose the width is shorter than the interaction range, namely $W < \lambda (\ll L)$, the Coulomb interaction between the edge channels results in $g_0 \propto W^{-1}$.  However, if the screening of the Coulomb interaction is relevant, $g_0$ would decay exponentially as $W$ increases.  
Even when the two-dimensional bulk states are highly disordered with a glassy behavior, screening according to the Thomas--Fermi theory might occur \cite{Lee-Ramakrishnan}.  
Regarding the configuration of the edge channels, when we take account of the electrostatic potential, the edge states may turn into compressible strips \cite{Chklovskii}, which may also affect the screening.  
The suppression of electron-electron interaction and hence small $g_0$ weaken the nonlinear response.  It is actually desirable for the use of the quantum Hall effect as resistance standard, where the typical device width is hundreds of micrometers \cite{Jeckelmann,Poirier}.  
On the other hand, much narrower devices with larger $g_0$ are favorable to observe the nonlinear response.

While we may arbitrarily change $\Phi_\text{L}$ and $\Phi_\text{R}$ separately, we can fix their relation, for example, by imposing the particle number conservation in the system.  Under the condition, the system equilibrates to be the original equilibrium state when the edge states are shunted and the system is coupled to a thermal bath without particle transfer.  
The condition $N_\text{L} + N_\text{R} = (n_\text{L} + n_\text{R})L = 0$ with Eqs.~\eqref{eq:density}, \eqref{eq:self-consistent}, \eqref{eq:voltage} results in 
\begin{align}
\label{eq:voltage_condition}
\Phi_p = -p \frac{v_p}{v_\text{L} + v_\text{R}} V_\text{H} - \frac{v_\text{L} v'_\text{R} + v_\text{R} v'_\text{L}}{(v_\text{L} + v_\text{R})^3} V_\text{H}^2 + O(g_0, V_\text{H}^3), 
\end{align}
which describes the the voltage $\Phi_p$ as a function of the Hall voltage $V_\text{H} = \Phi_\text{R} - \Phi_\text{L}$.  
Then, Eq.~\eqref{eq:current-response} leads to the current response $I$ in terms of the Hall voltage $V_\text{H}$: 
\begin{align}
\label{eq:current-response_condition}
I &= \frac{e^2}{h} V_\text{H} 
+ \frac{2g_0 e^3 (v'_\text{L} - v'_\text{R})}{h^2 (v_\text{L} + v_\text{R})^3} V_\text{H}^2 \nonumber\\
&\quad - \frac{2 g_0 e^4 (v_\text{L} - v_\text{R}) (v'_\text{L} - v'_\text{R}) (v_\text{L} v'_\text{R} + v_\text{R} v'_\text{L})}{h^2 (v_\text{L} + v_\text{R})^5 v_\text{L} v_\text{R}} V_\text{H}^3 \nonumber\\
&\quad + \frac{2 g_0 e^4 (v'_\text{L} - v'_\text{R}) (v_\text{L} v'_\text{R} - v_\text{R} v'_\text{L})}{h^2 (v_\text{L} + v_\text{R})^4 v_\text{L} v_\text{R}} V_\text{H}^3 
+ O(g_0^2, V_\text{H}^4).
\end{align}

\begin{figure}
\centering
\includegraphics[width=\hsize]{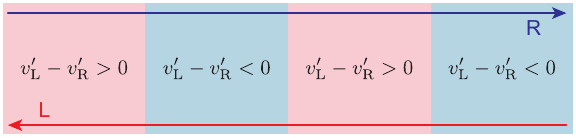}
\caption{Inhomogeneous system with varying energy dispersion curvatures.  
The difference $v'_\text{L} - v'_\text{R}$ is positional dependent.  We may regard the system as an array of quantum Hall systems with different values of $v'_\text{L} - v'_\text{R}$.  
}
\label{fig:disorder}
\end{figure}

To highlight the effects of the curvature of the energy dispersion and the interaction, we put $v_\text{L} = v_\text{R} \equiv v$.  
Then, Eq.~\eqref{eq:current-response_condition} becomes 
\begin{align}
\label{eq:current-response-voltage}
I
&= \frac{e^2}{h} V_\text{H} 
+ \frac{g_0 e^3}{4h^2 v^3} (v'_\text{L} - v'_\text{R}) V_\text{H}^2
- \frac{g_0 e^4}{8h^2 v^5} (v'_\text{L} - v'_\text{R})^2 V_\text{H}^3 \nonumber\\
&\quad + O(g_0^2, V_\text{H}^4).  
\end{align}
In reality, the energy dispersion of a quantum Hall edge channel depends on the confining potential near the edge, the dielectric environment, etc., and thus it may vary in space.  The curvature of the energy dispersion along the edge would particularly be dependent on the edge environment whereas the sign of the velocity is constrained by the chirality of the edge state.  
If the edge transport channels are inhomogeneous and the curvatures of the energy dispersions vary in space, we may split the system into regions inside which $v'_\text{L}$ and $v'_\text{R}$ are constant (Fig.~\ref{fig:disorder}).  In addition, if the interaction is constrained within each region, i.e., the interaction range $\lambda$ is smaller than the length of the region, we can regard the system as an array of quantum Hall systems with random curvatures of the energy dispersion.  
We then consider spatial averaging of $v'_\text{L} - v'_\text{R}$ and $(v'_\text{L} - v'_\text{R})^2$, which appear as coefficients of the second- and third-order response in Eq.~\eqref{eq:current-response-voltage}, respectively.  
When the curvature $v'_p$ takes a random value for each region, the difference $v'_\text{L} - v'_\text{R}$ vanishes but its square $(v'_\text{L} - v'_\text{R})^2$ remains finite  on average.  As a result, the lowest-order nonlinear response appears at \textit{third order} in the Hall voltage $V_\text{H}$.  This is consistent with a symmetry consideration as inversion symmetry recovers after spatial averaging.

Though we have concentrated on a steady state throughout the analysis, we may extend the result to a finite frequency case when the frequency is low enough not to induce a transition between the Landau levels and the periodicity is longer than the transition time to a stationary state.  Then, we can separately measure the linear and third-order voltage responses using the lock-in technique as they appear at different frequencies.  
We note that Eq.~\eqref{eq:current-response_condition} describes \textit{in-phase} response without out-of-phase components, which may arise from capacitive or inductive or dissipative origins.   
In a Hall bar setup (Fig.~\ref{fig:setup}), we apply an oscillating current $I\sin(\omega t)$ flowing between the source and the drain, and measure the longitudinal and Hall voltages, $V_\text{L}$ and $V_\text{H}$, respectively.  
In such a case, in-phase responses refer to voltage responses proportional to $\sin(m \omega t)$ and out-of-phase responses to $\cos(m \omega t)$ with an integer $m$.

Lastly, we discuss the case where multiple chiral edge channels $(\nu > 1)$ reside along the edges when the higher Landau levels are occupied.  
While we leave an explicit calculation, we can argue how the nonlinear response depends on the filling factor $\nu$ as follows.  
For the linear quantum Hall response, the conductance simply counts the number of chiral edge channels, and thus the linear Hall conductance $G_\text{H}$ is proportional to $\nu$: $G_\text{H} = \nu e^2/h$.  
For the nonlinear response, on the other hand, electron-electron interaction couples every pair of counterpropagating edge channels located on the opposite side of the sample.  Therefore, the nonlinear current response should be proportional to $\nu^2$, reflecting the nature of the two-body interaction.

\section{Materials and Methods}

To identify the unitary transformation $U$, in hindsight, it is easier to perform the unitary transformation $U^{-1} = U^\dagger$ on the noninteracting Hamiltonian $H_0$ to find the interacting Hamiltonian $H_\text{ex}$ from $U H_0 U^\dagger$, as opposed to Eq.~\eqref{eq:unitary-mapping}.  
We now show that the unitary operator $U$ that implements the mapping $U^\dagger: H_0 \mapsto H_\text{ex}$ and hence Eq.~\eqref{eq:unitary-mapping} is 
\begin{equation}
\label{eq:unitary}
U = e^{\Omega}, \quad 
\Omega = -\frac{1}{2} \sum_{q \neq 0} \sum_p p \frac{\alpha_q}{\eta_q} \tilde{\rho}_{p -q} \tilde{\rho}_{-p q}, 
\end{equation}
where $\alpha_q$ is an even function of $q$ $(\alpha_q = \alpha_{-q})$ and $\Omega$ is an anti-Hermitian operator $(\Omega^\dagger = -\Omega)$.  
The identification of $\alpha_q$, which we perform in the following, completes the mapping.  
It is worth noting that the unitary transformation $U$ does not change the zero modes, i.e., the Fourier components with the wavenumber zero.  
The operator $U$ transforms the density operator $\tilde{\rho}_{pq}$ as 
\begin{equation}
\label{eq:unitary_expression}
\begin{gathered}
U \tilde{\rho}_{pq} U^\dagger = u_q \tilde{\rho}_{pq} + v_q \tilde{\rho}_{-p q} \quad (q \neq 0), \\
U N_p U^\dagger = N_p \quad (N_p = \tilde{\rho}_{pq}|_{q=0}), 
\end{gathered}
\end{equation}
where we write $u_q = \cosh\alpha_q$ and $v_q = \sinh\alpha_q$.  
Since the zero modes $N_p$ are invariant under the unitary transformation, the real-space representation is 
\begin{equation}
\label{eq:unitary-rho}
U \tilde{\rho}_p(x) U^\dagger = \tilde{\rho}_p^u(x) + \tilde{\rho}_p^v(x) + \delta\tilde{\rho}_{p0}. 
\end{equation}
Here we introduce the notations 
\begin{gather}
\displaybreak[0]
\label{eq:rho-uv}
\tilde{\rho}_p^u(x) = (u \ast \tilde{\rho}_p)(x), \quad
\tilde{\rho}_p^v(x) = (v \ast \tilde{\rho}_p)(x), \\
\label{eq:delta-rho}
\delta\tilde{\rho}_{p0} = \frac{1}{L} [ (1-u_0)N_p - v_0 N_{-p} ], 
\end{gather}
with the convolution $(f \ast g)(x) = \int dy\, f(y) g(x-y)$.  $u(x)$ and $v(x)$ are the inverse Fourier transformations of $u_q$ and $v_q$, respectively.  

$\rho_{pq}$ and $\tilde{\rho}_{pq}$ distinguish the operators before and after the unitary transformation $U$.  From Eq.~\eqref{eq:unitary_expression}, we find the relation  
\begin{align}
\tilde{\rho}_{pq} = 
\begin{cases}
u_q \rho_{pq} + v_q \rho_{-p q} & (q \neq 0), \\
\rho_{pq} & (q = 0). 
\end{cases}
\end{align}
The substitution of $\tilde{\rho}_{pq}$ in $H_0$ implements the unitary transformation $U^\dagger: H_0 \mapsto H_\text{ex}$.  From the noninteracting Hamiltonian Eq.~\eqref{eq:H0_bosonized}, we obtain the interacting Hamiltonian Eq.~\eqref{eq:Hex} with the mapping 
\begin{subequations}
\label{eq:mapping2}
\begin{gather}
\label{eq:v}
v_p = u_0^2 (\tilde{v}_p + 4\pi\tilde{v}'_p \delta\tilde{\rho}_{p0}) + v_0^2 (\tilde{v}_{-p} + 4\pi\tilde{v}'_{-p} \delta\tilde{\rho}_{-p0}), \\
\label{eq:v-prime}
v'_p = u_0^3 \tilde{v}'_p + v_0^3 \tilde{v}'_{-p}, \\
\label{eq:g}
g(x-x') = \sum_p (2\pi \tilde{v}_p + 8\pi^2 \tilde{v}'_p \delta\tilde{\rho}_{p0}) (u \ast v)(x-x'), \\
\label{eq:g-prime}
g'_p = 2\pi (\tilde{v}'_p u_0^2 v_0 + \tilde{v}'_{-p} u_0 v_0^2), \\
\mu_p = \tilde{\mu}_p.  
\end{gather}
\end{subequations}
Here, we neglect irrelevant terms in the renormalization-group sense, i.e., higher-order terms in $\rho_p(x)$ or terms with higher-order derivatives as well as a constant shift of the chemical potential, which does not affect the ground state.  
The latter is simply a shift of the origin in measuring the energy, and thus it does not affect transport properties.  
We now confirm the existence of the unitary transformation Eq.~\eqref{eq:unitary-mapping}.  
By solving Eq.~\eqref{eq:mapping2} for the fictitious parameters, we obtain Eq.~\eqref{eq:mapping}.  

We note that the corrections from the zero-mode contribution $\delta\tilde{\rho}_{p0}$ appear at a finite density $N_p/L$.  
When the energy dispersion has a quadratic term in the wavenumber $k$, the Fermi velocity changes with the Fermi level even in a noninteracting case.  
As we are interested in the nonlinear transport, we retain all finite-density contributions, i.e., terms with $\delta\tilde{\rho}_{p0}$ in the following nonlinear transport analysis.  
Furthermore, keeping $\delta\tilde{\rho}_{p0}$ ensures the particle number conservation in the system, which we can confirm from Eq.~\eqref{eq:unitary-rho}.

\acknowledgments
We thank A. Rosch and A. H. MacDonald for useful discussion. 
This work was supported by JSPS KAKENHI Grant Number JP24H00197. 
H.I. was supported by Tokuo Fujii Research Fund. 
N.N. was supported by the RIKEN TRIP initiative and JSPS KAKENHI Grant Number 24H02231.


\begin{thebibliography}{99}

\bibitem{transistor}
J. Bardeen and W. H. Brattain, The Transistor, A Semi-Conductor Triode, Phys. Rev. \textbf{74}, 230 (1948).

\bibitem{Shockley}
W. Shockley, The Theory of p-n Junctions in Semiconductors and p-n Junction Transistors, Bell Syst. Tech. J. \textbf{28}, 435 (1949).

\bibitem{Tokura-Nagaosa}
Y. Tokura and N. Nagaosa, Nonreciprocal responses from non-centrosymmetric quantum materials, Nat. Commun. \textbf{9}, 3740 (2018).

\bibitem{Ono}
F. Ando, Y. Miyasaka, T. Li, J. Ishizuka, T. Arakawa, Y. Shiota, T. Moriyama, Y. Yanase, and T. Ono, Observation of superconducting diode effect, Nature \textbf{584}, 373 (2020).

\bibitem{Bauriedl}
L. Bauriedl, C. Bäuml, L. Fuchs, C. Baumgartner, N. Paulik, J. M. Bauer, K.-Q. Lin, J. M. Lupton, T. Taniguchi, K. Watanabe, C. Strunk, and N. Paradiso, Supercurrent diode effect and magnetochiral anisotropy in few-layer NbSe$_2$, Nat. Commun. \textbf{13}, 4266 (2022).

\bibitem{Daido}
A. Daido, Y. Ikeda, and Y. Yanase, Intrinsic Superconducting Diode Effect, Phys. Rev. Lett. \textbf{128}, 037001 (2022).

\bibitem{Yuan}
N. F. Q. Yuan and L. Fu, Supercurrent diode effect and finite-momentum superconductors, Proc. Natl. Acad. Sci. U.S.A. \textbf{119}, e2119548119 (2022).

\bibitem{JHe}
J. J. He, Y. Tanaka, and N. Nagaosa, A phenomenological theory of superconductor diodes, New J. Phys. \textbf{24}, 053014 (2022).

\bibitem{Tanaka}
Y. Tanaka, B. Lu, and N. Nagaosa, Theory of giant diode effect in $d$-wave superconductor junctions on the surface of a topological insulator, Phys. Rev. B \textbf{106}, 214524 (2022).

\bibitem{Lu}
B. Lu, S. Ikegaya, P. Burset, Y. Tanaka, and N. Nagaosa, Tunable Josephson Diode Effect on the Surface of Topological Insulators, Phys. Rev. Lett. \textbf{131}, 096001 (2023).

\bibitem{JHe2}
J. J. He, Y. Tanaka, and N. Nagaosa, The supercurrent diode effect and nonreciprocal paraconductivity due to the chiral structure of nanotubes, Nat. Commun. \textbf{14}, 3330 (2023).

\bibitem{Hu}
J. Hu, C. Wu, and X. Dai, Proposed Design of a Josephson Diode, Phys. Rev. Lett. \textbf{99}, 067004 (2007).

\bibitem{Misaki}
K. Misaki and N. Nagaosa, Theory of the nonreciprocal Josephson effect, Phys. Rev. B \textbf{103}, 245302 (2021).

\bibitem{Baumgartner}
C. Baumgartner, L. Fuchs, A. Costa, S. Reinhardt, S. Gronin, G. C. Gardner, T. Lindemann, M. J. Manfra, P. E. Faria Junior, D. Kochan, J. Fabian, N. Paradiso, and C. Strunk, Supercurrent rectification and magnetochiral effects in symmetric Josephson junctions, Nat. Nanotechnol. \textbf{17}, 39 (2022).

\bibitem{Ali}
H. Wu, Y. Wang, Y. Xu, P. K. Sivakumar, C. Pasco, U. Filippozzi, S. S. P. Parkin, Y.-J. Zeng, T. McQueen, and M. N. Ali, The field-free Josephson diode in a van der Waals heterostructure, Nature \textbf{604}, 653 (2022).

\bibitem{Pal}
B. Pal, A. Chakraborty, P. K. Sivakumar, M. Davydova, A. K. Gopi, A. K. Pandeya, J. A. Krieger, Y. Zhang, M. Date, S. Ju, N. Yuan, N. B. M. Schröter, L. Fu, and S. S. P. Parkin, Josephson diode effect from Cooper pair momentum in a topological semimetal, Nat. Phys. \textbf{18}, 1228 (2022).

\bibitem{Trahms}
M. Trahms, L. Melischek, J. F. Steiner, B. Mahendru, I. Tamir, N. Bogdanoff, O. Peters, G. Reecht, C. B. Winkelmann, F. von Oppen, and K. J. Franke, Diode effect in Josephson junctions with a single magnetic atom, Nature \textbf{615}, 628 (2023).

\bibitem{Efetov}
J. Díez-Mérida, A. Díez-Carlón, S. Y. Yang, Y.-M. Xie, X.-J. Gao, J. Senior, K. Watanabe, T. Taniguchi, X. Lu, A. P. Higginbotham, K. T. Law, and D. K. Efetov, Symmetry-broken Josephson junctions and superconducting diodes in magic-angle twisted bilayer graphene, Nat. Commun. \textbf{14}, 2396 (2023).

\bibitem{King-Smith}
R. D. King-Smith and D. Vanderbilt, Theory of polarization of crystalline solids, Phys. Rev. B \textbf{47}, 1651 (1993).

\bibitem{Vanderbilt}
R. Resta and D. Vanderbilt, ``Theory of Polarization: A Modern Approach'' in Physics of Ferroelectrics (Springer Berlin Heidelberg, Berlin, Heidelberg, 2007), vol. 105 of Topics in Applied Physics, pp. 31–68.

\bibitem{Sodemann}
I. Sodemann and L. Fu, Quantum Nonlinear Hall Effect Induced by Berry Curvature Dipole in Time-Reversal Invariant Materials, Phys. Rev. Lett. \textbf{115}, 216806 (2015).

\bibitem{Ma}
Q. Ma, S.-Y. Xu, H. Shen, D. MacNeill, V. Fatemi, T.-R. Chang, A. M. Mier Valdivia, S. Wu, Z. Du, C.-H. Hsu, S. Fang, Q. D. Gibson, K. Watanabe, T. Taniguchi, R. J. Cava, E. Kaxiras, H.-Z. Lu, H. Lin, L. Fu, N. Gedik, and P. Jarillo-Herrero, Observation of the nonlinear Hall effect under time-reversal-symmetric conditions, Nature \textbf{565}, 337 (2018).

\bibitem{Mak}
K. Kang, T. Li, E. Sohn, J. Shan, and K. F. Mak, Nonlinear anomalous Hall effect in few-layer WTe$_2$, Nat. Mater. \textbf{18}, 324 (2019).

\bibitem{Du-review}
Z. Z. Du, H.-Z. Lu, and X. C. Xie, Nonlinear Hall effects, Nat. Rev. Phys. \textbf{3}, 744 (2021).

\bibitem{Tiwari}
A. Tiwari, F. Chen, S. Zhong, E. Drueke, J. Koo, A. Kaczmarek, C. Xiao, J. Gao, X. Luo, Q. Niu, Y. Sun, B. Yan, L. Zhao, and A. W. Tsen, Giant c-axis nonlinear anomalous Hall effect in Td-MoTe$_2$ and WTe$_2$, Nat. Commun. \textbf{12}, 2049 (2021).

\bibitem{Lai}
S. Lai, H. Liu, Z. Zhang, J. Zhao, X. Feng, N. Wang, C. Tang, Y. Liu, K. S. Novoselov, S. A. Yang, and W. Gao, Third-order nonlinear Hall effect induced by the Berry-connection polarizability tensor, Nat. Nanotechnol. \textbf{16}, 869 (2021).

\bibitem{Yao}
J. Duan, Y. Jian, Y. Gao, H. Peng, J. Zhong, Q. Feng, J. Mao, and Y. Yao, Giant Second-Order Nonlinear Hall Effect in Twisted Bilayer Graphene, Phys. Rev. Lett. \textbf{129}, 186801 (2022).

\bibitem{Jack}
S. Sankar, R. Liu, X.-J. Gao, Q.-F. Li, C. Chen, C.-P. Zhang, J. Zheng, Y.-H. Lin, K. Qian, R.-P. Yu, X. Zhang, Z. Y. Meng, K. T. Law, Q. Shao, and B. Jäck, Observation of the Berry curvature quadrupole induced nonlinear anomalous Hall effect at room temperature, arXiv:2303.03274. 

\bibitem{high-frequency}
H. Isobe, S.-Y. Xu, and L. Fu, High-frequency rectification via chiral Bloch electrons, Sci. Adv. \textbf{6}, eaay2497 (2020).

\bibitem{Du}
Z. Z. Du, C. M. Wang, S. Li, H.-Z. Lu, and X. C. Xie, Disorder-induced nonlinear Hall effect with time-reversal symmetry, Nat. Commun. \textbf{10}, 3047 (2019).

\bibitem{Nandy}
S. Nandy and I. Sodemann, Symmetry and quantum kinetics of the nonlinear Hall effect, Phys. Rev. B \textbf{100}, 195117 (2019).

\bibitem{He1}
P. He, H. Isobe, D. Zhu, C.-H. Hsu, L. Fu, and H. Yang, Quantum frequency doubling in the topological insulator Bi$_2$Se$_3$, Nat. Commun. \textbf{12}, 698 (2021).

\bibitem{He2}
P. He, G. K. W. Koon, H. Isobe, J. Y. Tan, J. Hu, A. H. C. Neto, L. Fu, and H. Yang, Graphene moiré superlattices with giant quantum nonlinearity of chiral Bloch electrons, Nat. Nanotechnol. \textbf{17}, 378 (2022).

\bibitem{Ma-review}
Q. Ma, A. G. Grushin, and K. S. Burch, Topology and geometry under the nonlinear electromagnetic spotlight, Nat. Mater. \textbf{20}, 1601 (2021).

\bibitem{QHE}
R. E. Prange and S. M. Girvin, Eds., The Quantum Hall Effect (Springer US, New York, NY, 1987).

\bibitem{Klitzing}
K. v. Klitzing, G. Dorda, and M. Pepper, New Method for High-Accuracy Determination of the Fine-Structure Constant Based on Quantized Hall Resistance, Phys. Rev. Lett. \textbf{45}, 494 (1980).

\bibitem{Ando_QHE}
T. Ando, Y. Matsumoto, and Y. Uemura, Theory of Hall Effect in a Two-Dimensional Electron System, J. Phys. Soc. Jpn. \textbf{39}, 279 (1975).

\bibitem{Wakabayashi}
J. Wakabayashi and S. Kawaji, Hall Effect in Silicon MOS Inversion Layers under Strong Magnetic Fields, J. Phys. Soc. Jpn. \textbf{44}, 1839 (1978).

\bibitem{AHE} 
N. Nagaosa, J. Sinova, S. Onoda, A. H. MacDonald, and N. P. Ong, Anomalous Hall effect, Rev. Mod. Phys. \textbf{82}, 1539 (2010).

\bibitem{wavepacket}
D. Xiao, M.-C. Chang, and Q. Niu, Berry phase effects on electronic properties, Rev. Mod. Phys. \textbf{82}, 1959 (2010).

\bibitem{SHE}
J. Sinova, S. O. Valenzuela, J. Wunderlich, C. H. Back, and T. Jungwirth, Spin Hall effects, Rev. Mod. Phys. \textbf{87}, 1213 (2015).

\bibitem{Bhalla}
P. Bhalla, M.-X. Deng, R.-Q. Wang, L. Wang, and D. Culcer, Nonlinear Ballistic Response of Quantum Spin Hall Edge States, Phys. Rev. Lett. \textbf{127}, 206801 (2021).

\bibitem{Wang}
Z. Wang, P. Bhalla, M. Edmonds, M. S. Fuhrer, and D. Culcer, Unidirectional magnetotransport of linearly dispersing topological edge states, Phys. Rev. B \textbf{104}, L081406 (2021).

\bibitem{Laughlin}
R. B. Laughlin, Quantized Hall conductivity in two dimensions, Phys. Rev. B \textbf{23}, 5632 (1981).

\bibitem{Thouless}
D. J. Thouless, Localisation and the two-dimensional Hall effect, J. Phys. C \textbf{14}, 3475 (1981).

\bibitem{Prange}
R. E. Prange, Quantized Hall resistance and the measurement of the fine-structure constant, Phys. Rev. B \textbf{23}, 4802 (1981).

\bibitem{Aoki}
H. Aoki and T. Ando, Effect of localization on the hall conductivity in the two-dimensional system in strong magnetic fields, Solid State Commun. \textbf{38}, 1079 (1981).

\bibitem{TKNN}
D. J. Thouless, M. Kohmoto, M. P. Nightingale, and M. den Nijs, Quantized Hall Conductance in a Two-Dimensional Periodic Potential, Phys. Rev. Lett. \textbf{49}, 405 (1982).

\bibitem{CODATA}
E. Tiesinga, P. J. Mohr, D. B. Newell, and B. N. Taylor, CODATA recommended values of the fundamental physical constants: 2018, Rev. Mod. Phys. \textbf{93}, 025010 (2021).

\bibitem{Jeckelmann}
B. Jeckelmann and B. Jeanneret, The quantum Hall effect as an electrical resistance standard, Rep. Prog. Phys. \textbf{64}, 1603 (2001).

\bibitem{Poirier}
W. Poirier and F. Schopfer, Resistance metrology based on the quantum Hall effect, Eur. Phys. J. Spec. Top. \textbf{172}, 207 (2009).

\bibitem{Kohmoto}
M. Kohmoto, Topological invariant and the quantization of the Hall conductance, Ann. Phys. \textbf{160}, 343 (1985).

\bibitem{Niu}
Q. Niu, D. J. Thouless, and Y.-S. Wu, Quantized Hall conductance as a topological invariant, Phys. Rev. B \textbf{31}, 3372 (1985).

\bibitem{Cage1}
M. E. Cage, R. F. Dziuba, B. F. Field, E. R. Williams, S. M. Girvin, A. C. Gossard, D. C. Tsui, and R. J. Wagner, Dissipation and Dynamic Nonlinear Behavior in the Quantum Hall Regime, Phys. Rev. Lett. \textbf{51}, 1374 (1983).

\bibitem{Kawaji}
S. Kawaji, K. Hirakawa, M. Nagata, T. Okamoto, T. Fukuse, and T. Gotoh, Breakdown of the Quantum Hall Effect in GaAs/AlGaAs Heterostructures Due to Current, J. Phys. Soc. Jpn. \textbf{63}, 2303 (1994).

\bibitem{Tsemekhman}
V. Tsemekhman, K. Tsemekhman, C. Wexler, J. H. Han, and D. J. Thouless, Theory of the breakdown of the quantum Hall effect, Phys. Rev. B \textbf{55}, R10201 (1997).

\bibitem{Nachtwei}
G. Nachtwei, Breakdown of the quantum Hall effect, Physica E \textbf{4}, 79 (1999).

\bibitem{Alexander-Webber}
J. A. Alexander-Webber, A. M. R. Baker, T. J. B. M. Janssen, A. Tzalenchuk, S. Lara-Avila, S. Kubatkin, R. Yakimova, B. A. Piot, D. K. Maude, and R. J. Nicholas, Phase Space for the Breakdown of the Quantum Hall Effect in Epitaxial Graphene, Phys. Rev. Lett. \textbf{111}, 096601 (2013).

\bibitem{Halperin}
B. I. Halperin, Quantized Hall conductance, current-carrying edge states, and the existence of extended states in a two-dimensional disordered potential, Phys. Rev. B \textbf{25}, 2185 (1982).

\bibitem{Oreg1}
Y. Oreg and A. M. Finkel’stein, Interedge Interaction in the Quantum Hall Effect, Phys. Rev. Lett. \textbf{74}, 3668 (1995).

\bibitem{Oreg2}
Y. Oreg and A. M. Finkel’stein, dc transport in quantum wires, Phys. Rev. B \textbf{54}, R14265 (1996).

\bibitem{experiment}
P. He, H. Isobe, G. K. W. Koon, J. Y. Tan, J. Hu, J. Li, N. Nagaosa, and J. Shen, Third-order nonlinear Hall effect in a quantum Hall system, Nat. Nanotechnol. \textbf{19}, 1460 (2024). 

\bibitem{Tomonaga}
S.-i. Tomonaga, Remarks on Bloch’s Method of Sound Waves applied to Many-Fermion Problems, Prog. Theor. Phys. \textbf{5}, 544 (1950).

\bibitem{Luttinger}
J. M. Luttinger, An Exactly Soluble Model of a Many‐Fermion System, J. Math. Phys. \textbf{4}, 1154 (1963).

\bibitem{Haldane}
F. D. M. Haldane, “Luttinger liquid theory” of one-dimensional quantum fluids. I. Properties of the Luttinger model and their extension to the general 1D interacting spinless Fermi gas, J. Phys. C \textbf{14}, 2585 (1981).

\bibitem{Rozhkov1}
A. V. Rozhkov, Fermionic quasiparticle representation of Tomonaga-Luttinger Hamiltonian, Eur. Phys. J. B \textbf{47}, 193 (2005).

\bibitem{Rozhkov2}
A. V. Rozhkov, Class of exactly soluble models of one-dimensional spinless fermions and its application to the Tomonaga-Luttinger Hamiltonian with nonlinear dispersion, Phys. Rev. B \textbf{74}, 245123 (2006).

\bibitem{Rozhkov3}
A. V. Rozhkov, Density-density propagator for one-dimensional interacting spinless fermions with nonlinear dispersion and calculation of the Coulomb drag resistivity, Phys. Rev. B \textbf{77}, 125109 (2008).

\bibitem{Glazman2}
M. Pustilnik, M. Khodas, A. Kamenev, and L. Glazman, Dynamic Response of One-Dimensional Interacting Fermions, Phys. Rev. Lett. \textbf{96}, 196405 (2006).

\bibitem{Glazman3}
M. Khodas, M. Pustilnik, A. Kamenev, and L. I. Glazman, Fermi-Luttinger liquid: Spectral function of interacting one-dimensional fermions, Phys. Rev. B \textbf{76}, 155402 (2007).

\bibitem{Glazman4}
A. Imambekov and L. I. Glazman, Universal Theory of Nonlinear Luttinger Liquids, Science \textbf{323}, 228 (2009).

\bibitem{Glazman_review}
A. Imambekov, T. L. Schmidt, and L. I. Glazman, One-dimensional quantum liquids: Beyond the Luttinger liquid paradigm, Rev. Mod. Phys. \textbf{84}, 1253 (2012).

\bibitem{Mattis-Lieb}
D. C. Mattis and E. H. Lieb, Exact Solution of a Many-Fermion System and Its Associated Boson Field, J. Math. Phys. \textbf{6}, 304 (1965).

\bibitem{Kawabata}
A. Kawabata, On the Renormalization of Conductance in Tomonaga-Luttinger Liquid, J. Phys. Soc. Jpn. \textbf{65}, 30 (1996).

\bibitem{Kaplan}
D. Kaplan, T. Holder, and B. Yan, General nonlinear Hall current in magnetic insulators beyond the quantum anomalous Hall effect, Nat. Commun. \textbf{14}, 3053 (2023).

\bibitem{Hansson}
T. H. Hansson, M. Hermanns, S. H. Simon, and S. F. Viefers, Quantum Hall physics: Hierarchies and conformal field theory techniques, Rev. Mod. Phys. \textbf{89}, 025005 (2017).

\bibitem{Tarucha}
S. Tarucha, T. Honda, and T. Saku, Reduction of quantized conductance at low temperatures observed in 2 to 10 $\si{\um}$-long quantum wires, Solid State Commun. \textbf{94}, 413 (1995).

\bibitem{Maslov}
D. L. Maslov and M. Stone, Landauer conductance of Luttinger liquids with leads, Phys. Rev. B \textbf{52}, R5539 (1995).

\bibitem{Ponomarenko}
V. V. Ponomarenko, Renormalization of the one-dimensional conductance in the Luttinger-liquid model, Phys. Rev. B \textbf{52}, R8666 (1995).

\bibitem{Safi1}
I. Safi and H. J. Schulz, Transport in an inhomogeneous interacting one-dimensional system, Phys. Rev. B \textbf{52}, R17040 (1995).

\bibitem{Safi2}
I. Safi, Conductance of a quantum wire: Landauer’s approach versus the Kubo formula, Phys. Rev. B \textbf{55}, R7331 (1997).

\bibitem{Safi3}
I. Safi, A dynamic scattering approach for a gated interacting wire, Eur. Phys. J. B \textbf{12}, 451 (1999).

\bibitem{Shimizu}
A. Shimizu, Landauer Conductance and Nonequilibrium Noise of One-Dimensional Interacting Electron Systems, J. Phys. Soc. Jpn. \textbf{65}, 1162 (1996).

\bibitem{Lee-Ramakrishnan}
P. A. Lee and T. V. Ramakrishnan, Disordered electronic systems, Rev. Mod. Phys. \textbf{57}, 287–337 (1985).

\bibitem{Chklovskii}
D. B. Chklovskii, B. I. Shklovskii, and L. I. Glazman, Electrostatics of edge channels, Phys. Rev. B \textbf{46}, 4026–4034 (1992).


\end{thebibliography}
\end{document}